\documentclass[amsmath,amssymb,pra,aps,showpacs,superscriptaddress,twocolumn]{revtex4-2}
\usepackage{amsmath,amsfonts,amssymb,amsthm,graphics,graphicx,epsfig,bbm}
\usepackage[colorlinks=true,citecolor=blue,linkcolor=blue,urlcolor=blue]{hyperref}
\usepackage[usenames]{color}
\usepackage{graphicx}
\usepackage{subfigure}
\usepackage{amsmath}
\usepackage{epsfig}
\usepackage{dcolumn}
\usepackage{bm}
\usepackage{color}
\usepackage{times}
\usepackage{epstopdf}
\usepackage{amssymb}
\usepackage{amstext}
\usepackage{latexsym}
\usepackage{hyperref}
\usepackage{amsfonts}
\usepackage{psfrag}
\usepackage{soul,xcolor}
\usepackage[normalem]{ulem}
\usepackage{dsfont}
\usepackage{txfonts}
\usepackage{float}

\newcommand{\ket}[1]{\vert #1 \rangle}
\newcommand{\bra}[1]{\langle #1 \vert}
\newcommand{\ketbra}[2]{\vert #1 \rangle \langle #2 \vert}

\newcommand{\ie}{{\it{i.e.~}}}

\begin{document}

\setstcolor{red}

\title{Embedded Quantum Correlations in thermalized quantum Rabi systems}
\date{\today}

\author{M. Ahumada}
\affiliation{Departamento de F\'isica, Universidad de Santiago de Chile (USACH), Avenida V\'ictor Jara 3493, 9170124, Santiago, Chile}

\author{F. A. C\'ardenas-L\'opez}
\affiliation{Forschungszentrum J\"ulich GmbH, Peter Gr\"unberg Institute, Quantum Control (PGI-8), 52425 J\"ulich, Germany}

\author{G.  Alvarado Barrios}
\affiliation{Kipu Quantum, Greifswalderstrasse 226, 10405 Berlin, Germany}

\author{F. Albarr\'an-Arriagada}
\email[F. Albarr\'an-Arriagada]{\qquad francisco.albarran@usach.cl}
\affiliation{Departamento de F\'isica, Universidad de Santiago de Chile (USACH), Avenida V\'ictor Jara 3493, 9170124, Santiago, Chile}
\affiliation{Center for the Development of Nanoscience and Nanotechnology 9170124, Estaci\'on Central, Santiago, Chile}

\author{J. C. Retamal}
\affiliation{Departamento de F\'isica, Universidad de Santiago de Chile (USACH), Avenida V\'ictor Jara 3493, 9170124, Santiago, Chile}
\affiliation{Center for the Development of Nanoscience and Nanotechnology 9170124, Estaci\'on Central, Santiago, Chile}

\begin{abstract}
We study the quantum correlations embedded in open quantum Rabi systems. Specifically, we study how the quantum correlation depends on the coupling strength, number of qubits, and reservoir temperatures. We numerically calculate the quantum correlations of up to three qubits interacting with a single field mode. We find that the embedded quantum correlations exhibit a maximum for a given coupling strength,  which depends inversely on the number of subsystems and the reservoir temperature. We explore how this feature affects the performance of a many-qubit Otto heat engine, finding numerical evidence of a direct correspondence between the minimum of the extractable work and the maximum of the embedded quantum correlations in the qubit-cavity bi-partition. Furthermore, as we increase the number of qubits, the maximum extractable work is reached at smaller values of the coupling strength. This work could help design more sophisticated quantum heat engines that rely on many-body systems with embedded correlations as working substances.
\end{abstract}

\maketitle

\section{Introduction}

The study of quantum correlations embedded in multipartite quantum systems is a fundamental task in quantum information. Entanglement has been widely recognized as a central resource for the realization of quantum information and quantum computation applications. A prolific amount of research over many years led us to a deep understanding of entangled quantum systems, their characterization, and applications~\cite{Horodecki2009RevModPhys,Adesso2016JPA,Ekert1998PTRS}. In this sense, we have well-known measures of quantum entanglement, such as entanglement of formation, which measures the amount of entanglement of a bipartition in a quantum system~\cite{Wootters1998PRL,Bennett1996PRA}. There are other quantum correlations beyond entanglement to be scrutinized in bipartite and multipartite quantum systems~\cite{DeChiara2018RPP}. The calculation of all quantum correlations in a bipartite system can be done by subtracting the classical correlations from the mutual information, which quantifies all possible correlations, obtaining the well-known concept of quantum discord~\cite{Ollivier2001PRL, Henderson2001JPA}. The calculation of classical correlation, and therefore the calculation of quantum discord, involves an optimization process over all the possible projective measurements over one of the subsystems, which for a general system is a hard task~\cite{QDNPHuang}. 

Due to the complexity in the calculation of quantum correlations for high-dimension bipartite systems, its study has mainly focused on the two-qubit case~\cite{Ali2010PRA,Li2011PRA,Shi2011NJP,Shi2011JPA,LoFranco2012IJMPB}. Nevertheless, there have been efforts to evaluate quantum correlations in higher dimensions. For example, for the case of entanglement, there are proposals for bounds estimation~\cite{Gerjuoy2003PRA,Chen2005PRL,Bai2014PRL}, or numerical approaches attempting to solve the optimization process~\cite{Allende2015PRA,Toth2015PRL}. In the case of quantum discord in higher dimensions, the problem has been elusive, and to our knowledge, only a few special cases have been considered~\cite{Ma2015SciRep, Rau2018QIP, Chitambar2012PRA, Rossignoli2012PRA, Hou2013EPJ, Malavezzi2016PRB}. Then, the problem deserves attention and several possibilities to explore are still open. 

An important case that presents embedded quantum correlations in higher dimensions is the light-matter interaction beyond the strong coupling regime, described by the quantum Rabi model. The presence of rotating and counter-rotating interaction terms are responsible for the anharmonic spectrum and highly correlated eigenstates~\cite{Xie2017JPA,FornDiaz2019RMP,Rossatto2017PRA}. These properties have been studied in the context of quantum phase transitions~\cite{Cai2021NatCommun}, quantum metrology~\cite{Garbe2020PRL}, and quantum information processing~\cite{AlbarranArriagada2018PRA,Wang2016}. This model has also been extended to multimode and multiqubit generalizations, which present dark states in the energy spectrum allowing the development of applications for entangled state generation~\cite{Peng2021PRL}. Therefore, the quantum Rabi model and its variations present a good testbed to study the role and quantification of quantum correlations beyond entanglement for quantum technologies.

Quantum correlations embedded in eigenstates of multipartite quantum systems under the effect of a thermal environment could play an important role as resources for emerging quantum technologies that operate between stationary regimes as in the case of quantum heat engines (QHE). In recent years, QHE has received increasing attention, with fruitful experimental and theoretical progress~\cite{Quan2007PRE,Peterson2019PRL,Rossnagel2016Science,Abah2012PRL}. In this area, quantum Otto engines have been widely studied due to their simplicity and clear thermodynamical interpretation. In this context, embedded quantum correlations in the working substance may enhance the performance of the engine. Quantum correlations that are preserved in the thermal state of a quantum system may be related to the enhancement of the work extraction in QHE~\cite{Altintas2014PRE,Altintas2015PRA,Alvarado2017PRA}. Specifically, these systems have shown that the difference in the quantum correlations between the thermal states is related to the work extraction and efficiency of the engine. However, these efforts have been mainly focused on the case of highly entangled bipartite systems,  where the working substance is described by the single-qubit quantum Rabi model. Thus, it would be relevant to explore the generalization of such a hypothesis for highly correlated multipartite systems as working substances. 

In this work, we study a multipartite quantum system described by the multiqubit quantum Rabi model beyond the strong coupling. This system exhibits embedded quantum correlations in its eigenstates. We explore these quantum correlations in terms of the coupling strength for different temperatures and the number of qubits. We study the performance of a QHE and its relation with its embedded quantum correlations considering the multiqubit quantum Rabi model as the working substance.  

\begin{figure}[t]
\centering
\includegraphics[width=0.9\linewidth]{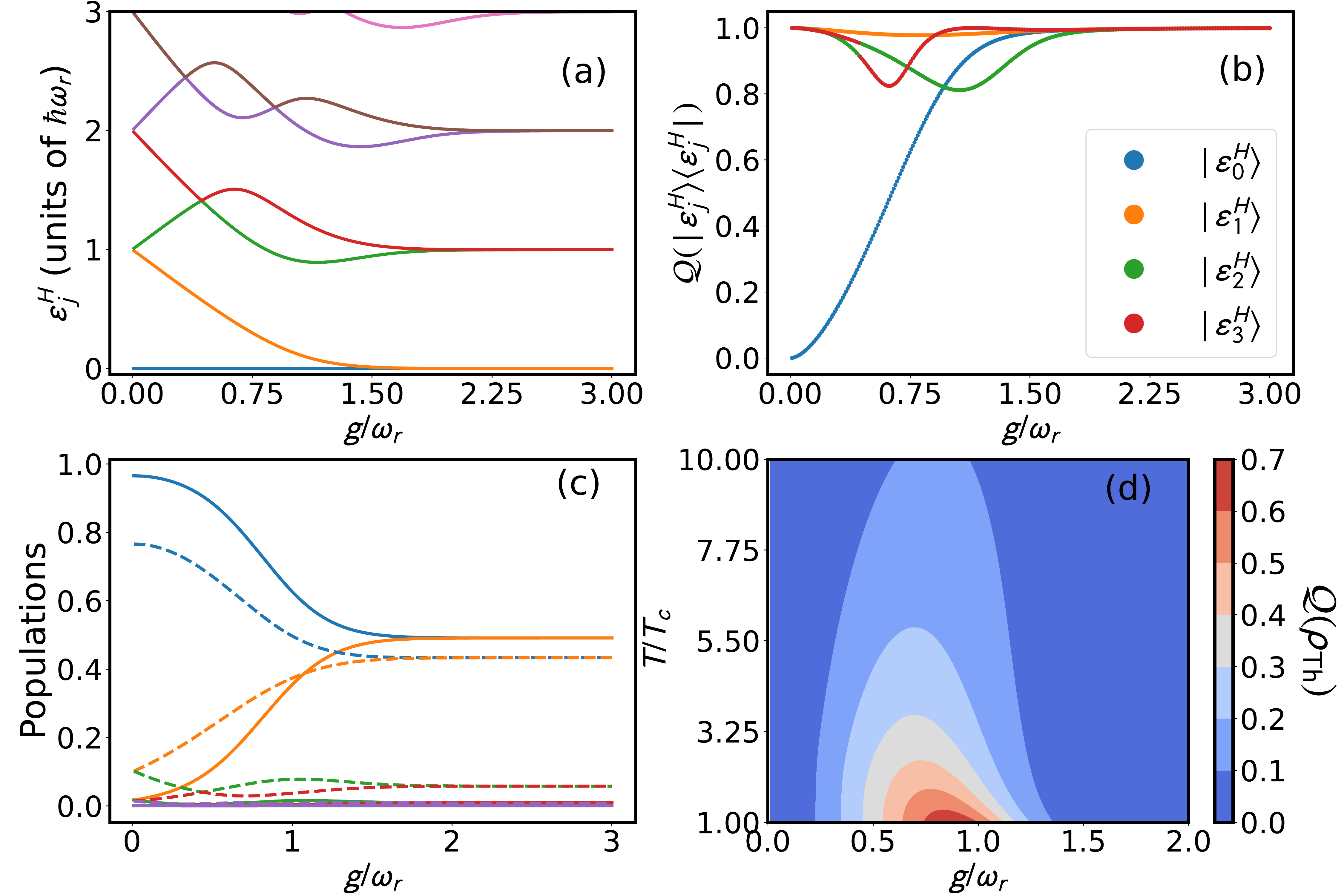}
\caption{The quantum Rabi model considering $N=1$ qubit: (a) Energy spectrum of the Hamiltonian $H$ as a function of the normalized coupling strength $g/\omega_{r}$. (b) Quantum correlations $\mathit{Q}(\ketbra{\epsilon_j^H}{\epsilon_j^H})$ of the first four energy states as a function of the normalized coupling strength $g/\omega_{r}$. (c) Thermal populations of the ground state and the first four excited states for $T=T_{c}$ (solid line) and $T=10T_{c}$ (dashed line) as a function of the normalized coupling strength $g/\omega_{r}$. (d) Contour plot: Quantum correlations of the thermal state $\mathit{Q}({\rho\textrm{th}})$ as a function of the normalized coupling strength $g/\omega_{r}$ and normalized temperature $T/T_{c}$. We have used $\omega_{q,l}=\omega_{r}=\omega$ and $\omega/2\pi=8\,\textrm{GHz}$, with $T_{c}=19\,\textrm{mK}$.}
\label{Fig01}
\end{figure}

\section{Model}

We consider a set of $N$ two-level systems (TLS) of frequency $\omega_{q,\ell}$ interacting with a quantized field mode of frequency $\omega_{r}$ described by the multiqubit quantum Rabi model, also called Dicke model~\cite{Garraway2011PTRSA,Li2019SciRep} which reads
\begin{equation}
H=\hbar \omega_{r} a^{\dagger}a+\hbar\sum_{\ell}^{N} \omega_{q,\ell} \sigma_{\ell}^{z} +\hbar \sum_{\ell}^{N}g_{\ell}\sigma^{x}_{\ell} (a+a^{\dagger}),
\label{Eq01}
\end{equation}
where $g_{\ell}$ stands for the coupling strength between the $\ell$th TLS with the field mode. The operator $\sigma^{x}_{\ell}$($\sigma^{z}_{\ell}$) correspond to the Pauli matrix $x$($z$) applied over the $\ell$th TLS. Finally,  $a(a^{\dag})$ is the annihilation (creation) bosonic operator for the field mode. In the strong coupling regime, $g/\omega_{r}\lesssim 0.1$, the model converges to the Tavis-Cumming model which preserves the number of excitations $N$, allowing to express the Hamiltonian with a finite $N$ dimensional subspace~\cite{TCmodel}. On the other hand, for larger coupling strength $0.1\lesssim g/\omega_{r}$ such symmetry breaks down, and the parity symmetry $\mathbb{Z}_{2}$ determines the internal Hamiltonian structure. This means that the Hamiltonian divides the Hilbert space into two orthogonal subspaces spanned by states with even/odd number of excitations, \ie, the eigenstates of the parity operator $\mathcal{P}=exp(-i\pi[\sum(\sigma_{\ell}^z+1)/2+a^{\dagger}a])$. Several theoretical proposals and experimental realizations have been done for the quantum Rabi model and its variants in different regimes, mainly in circuit quantum electrodynamics~\cite{Niemczyk2010NatPhys,Ballester2012PRX,Yoshihara2017NatPhys,Bosman2017npjQI,Blais2021RMP}. This model presents eigenstates with embedded multipartite quantum correlations which can be interesting to explore in thermal equilibrium.

We will focus on the thermal equilibrium of our multiqubit quantum Rabi system (MQRS) with a reservoir at temperature $T$, which density matrix is given by the thermal state
\begin{equation}
\rho_{\rm{th}}(T,H) =\sum_{j=1}^{\infty}P^{\textrm{th}}_j(T,H) \ketbra{\epsilon_j^H}{\epsilon_j^H}, 
\label{Eq02}
\end{equation}
where $\ket{\epsilon_j^H}$ is the eigenvector corresponding to the eigenenergy $\epsilon_j^H$ of the MQRS which Hamiltonian $H$ is given in Eq. (\ref{Eq01}). 
 We define $P^{\textrm{th}}_j(T,H)={e^{-{\epsilon_j^H}/{k_bT}}}/{Z(T,H)}$ as the thermal probability of the state $j$th state of $H$ at temperature $T$ , $Z(T,H) =\sum_j e^{-\epsilon_j^H/k_bT}$ is the partition function and $k_b$ is the Boltzmann constant.  We are interested in studying the correlations embedded in a thermal state of the MQRS as a function of the coupling strength $g_{\ell}$ and the temperature $T$ taking as reference a typical operational temperature achieved in superconducting circuits, \ie $T_c= 19~\rm{mK}$. We explore temperatures until ten times $T_c$, and different number of qubits, $N$, in the Hamiltonian given in Eq.~(\ref{Eq01}).  In all the discussions that follow we assume symmetric coupling of qubits to the field mode $g_\ell =g$.

\section{Quantum correlations and optimization}
Regarding quantum correlations, we consider the quantum discord (QD) embedded in a bipartition of a quantum system given by \cite{Ollivier2001PRL, Henderson2001JPA}
\begin{equation}
Q(\rho_{AB})=S(\rho_{B})-S(\rho_{AB})+\min_{\Pi^{B}} \left[S(\rho_{A}|\Pi_{j}^{B})\right],
\label{Eq03}
\end{equation}
where $S(\rho)=-\textrm{Tr}[\rho\log(\rho)]$ is the von Neumann entropy, $\rho_{B}=\textrm{Tr}_A[\rho_{AB}]$ is the reduced state of subsystem $B$. In our case, we consider as part $A$ the bosonic mode and as part $B$ the set of TLS.  $\Pi^B=\{\Pi_j^B\}$ is a set of projection operators describing von Neumann measurement over the system $B$, and $S(\rho_{A}|\Pi_{j}^{B})$ is the conditional entropy of subsystem $A$ after a measurement ${\Pi_{j}^{B}}$ on system $B$. Such conditional entropy is defined as  $S(\rho_{A}|\Pi_{j}^{B})=\sum_{j}p_jS(\rho_j)$ with $\rho_j=\Pi_{j}^{B}\rho_{AB}\Pi_{j}^{B}/p_{j}$.  It is necessary to mention that von Neumann type measurements can be written as  $\Pi_j^B=\mathbb{I}_A\otimes\ket{\phi_j}_B\bra{\phi_j}$, where the set of states $\{\ket{\phi_j}_B\}$ is an orthonormal basis for the subsystem $B$.  The minimization in Eq.~(\ref{Eq03}) is over all possible von Neumann type measurements, which can be found by considering an arbitrary unitary transformation over the states $\ket{\phi_j}_B$\cite{Luo2008PRA,Ali2010PRA}, such as
\begin{equation}
\Pi_j^B(\vec{\theta})=\mathbb{I}_A\otimes V(\vec{\theta})\ket{\phi_j}_B\bra{\phi_j}V^{\dagger}(\vec{\theta}).
\end{equation}
Then, the minimization in Eq.~(\ref{Eq03}) can be done by the optimization of the set of independent parameters $\vec{\theta}$ for a general unitary operation $V(\vec{\theta})$. It is known that such a unitary operator can be decomposed into a product of unitary operators acting on two-dimensional subspaces~\cite{NielsenChuang}, such that the number of independent parameters for an arbitrary $d$-dimensional unitary operator is $3d(d-1)/2$.  The unitary $V$ can be conveniently represented as follows 
\begin{equation}
V=\prod_{k=1}^{d-1}\prod_{n=1}^{d-k}V_{k,n},
\end{equation}
where the matrix $V_{k,n}$ reads
\begin{equation}
V_{k,n}=\left(\begin{array}{cccccccccc}
1&0& ..& & & & && &\\
0& 1& & & & & & & &\\
..&&..&&&&&&\\
&&&v_{k,k}&&&v_{k,k+n}&&&\\
&&&&..&&&&&\\
&&&&&..&&&&\\
&&&v_{k+n,k}&&&v_{k+n,k+n}&&&\\
&&&&&&&..&&..\\
&&&&&&&&1&0\\
&&&&&&&..&0&1\\
\end{array}\right),
\end{equation}
where $k=1,2,...,d-1$ and $n=1,2,...,d-k$ and we parametrize each matrix  $v_{k,n}$ in terms of three arbitrary parameters such that  $v_{k,k}=\sin ({\phi_{1}})e^{i\phi_{2}}$, $v_{k+n,k}=\cos (\phi_{1})e^{i\phi_{3}}$, $v_{k,k+n}=\cos (\phi_{1})e^{-i\phi_{3}}$ and $v_{k+n,k+n}=-\sin ({\phi_{1}})e^{-i\phi_{2}}$\cite{NielsenChuang}. As an example for a bipartition of size $N \otimes 8$ the general projective measurement acting on the second subspace can be parametrized in terms of 84 angles. 

In this work, we carried out the optimization using the python package scipy.optimize. In particular, we have used basin-hopping~\cite{basinhopping} with the conjugate gradient algorithm~\cite{Gradient_descen}. Basin-hopping is a method that allows the optimization process to escape from local minima and therefore be a reliable tool for global optimization.

\section{Numerical results}

\begin{figure}[t]
\centering
\includegraphics[width=0.9\linewidth]{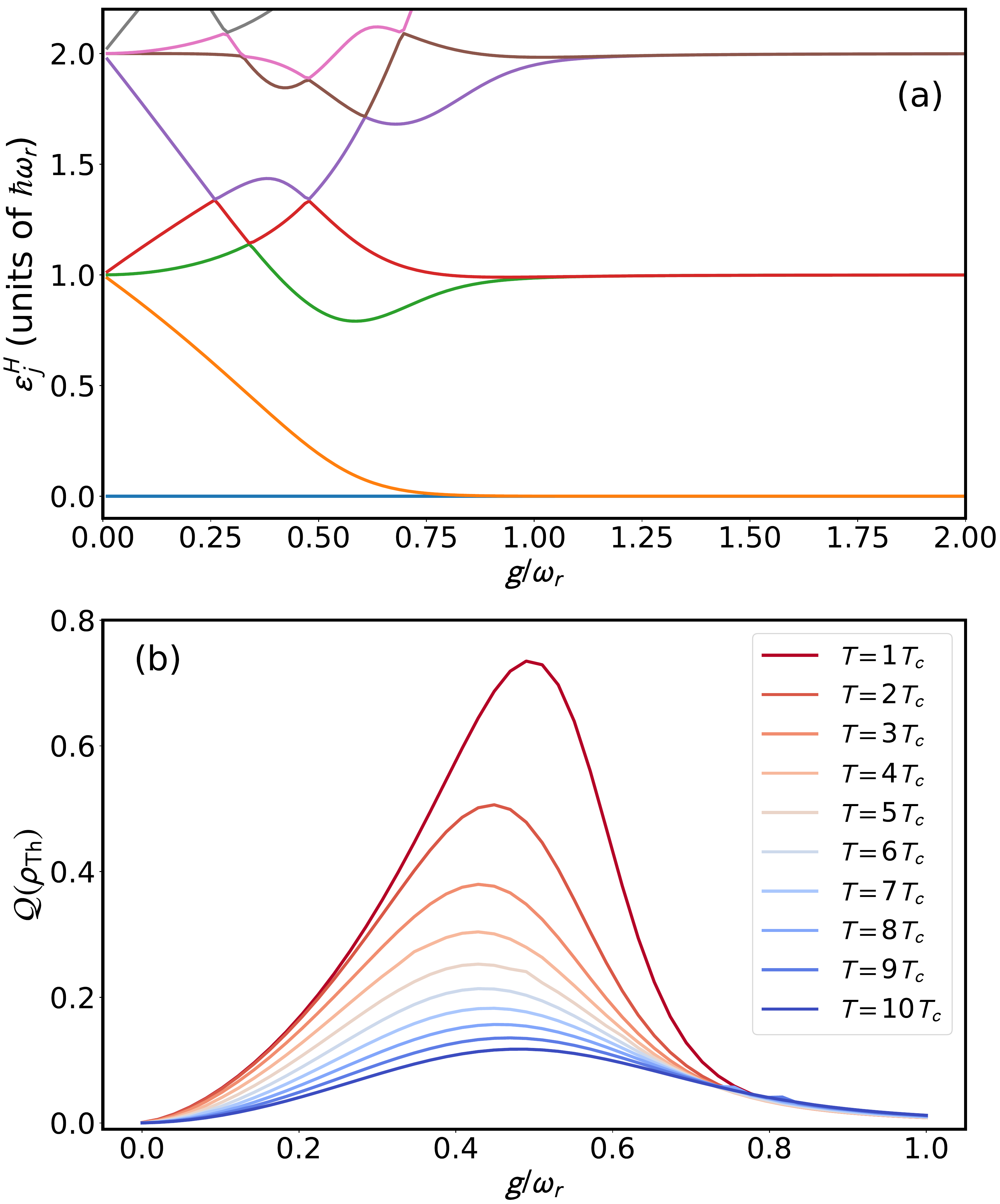}
\caption{The quantum Rabi model properties for $N=2$ qubits: 
(a) Energy spectrum of the Hamiltonian $H$ (referred to the ground state energy) as a function of the normalized coupling strength $g/\omega_{r}$. (b) Quantum correlations of the thermal state $\mathit{Q}({\rho\textrm{Th}})$ as a function of the normalized coupling strength $g/\omega_{r}$ for several temperatures. The temperatures of the thermal state are indicated in the panel. 
We have used $\omega_{q,l}=\omega_{r}=\omega$ and $\omega/2\pi=8\,\textrm{GHz}$, with $T_{c}=19\,\textrm{mK}$.}
\label{Fig02}
\end{figure}

\begin{figure*}[t]
\centering
\includegraphics[width=0.8\linewidth]{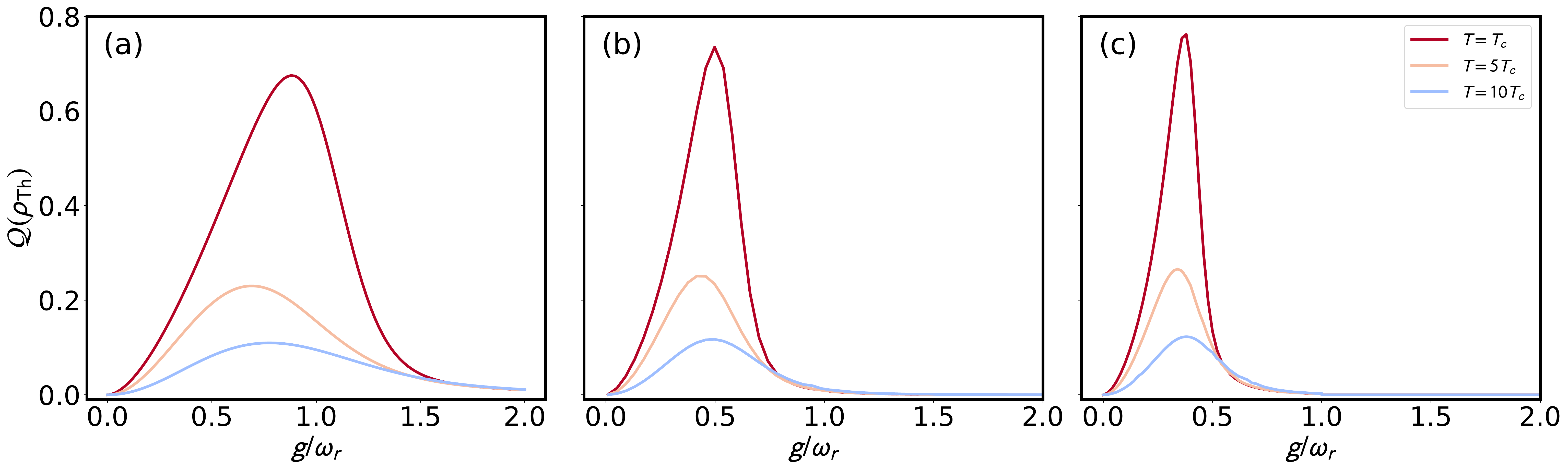}
\caption{The multi-qubit quantum Rabi model: The quantum correlations of the thermal state $\mathit{Q}(\rho_{\textrm{Th}})$ for (a) $N=1$, (b) $N=2$, and (c) $N=3$ qubits as a function of the normalized coupling strength $g/\omega_{r}$. The temperatures for the three cases  are indicated in the panel. We have used $\omega_{q,l}=\omega_{r}=\omega$ and $\omega/2\pi=8\,\textrm{GHz}$, with $T_{c}=19\,\textrm{mK}$.}
\label{Fig03}
\end{figure*} 

First, let us consider the standard quantum Rabi model, corresponding to $N=1$ in the Hamiltonian of Eq.~(\ref{Eq01}). Figure~\ref{Fig01}~(a) shows energy spectrum for this case (referred to the ground state energy) as a function of the normalized coupling strength $g/\omega_{r}$. In this case, the optimization is carried out over a unitary $V$ involving $3$ parameters, where Fig.~\ref{Fig01}~(b) shows the QD embedded in the first four energy states. Figure~\ref{Fig01}~(c) shows the population of the different eigenstates in a thermal state with temperature $T=T_c$ (solid line) and $T=10T_c$ (dashed line). We note that in both cases, the five lower energy states are enough to describe the system at such temperatures. Finally, Fig.~\ref{Fig01}~(d) shows the QD embedded in the thermal state as a function of the normalized temperature $T/T_c$ and the normalized coupling strength $g/\omega_r$.

At small $g/\omega_{r}$ the Rabi model corresponds to the well-known Jaynes-Cumming (JC) model, whose ground state is the vacuum, $\mid g,0\rangle$, and the excited states are given by $\mid \Psi_n^{\pm} \rangle =(\mid e, n \rangle \pm \mid g, n+1\rangle)/\sqrt{2}$. The excited states contain maximal correlations, as can be seen in Fig.~\ref{Fig01}~{(b)}, where at $g/\omega_{r}$  close to zero the correlation of the excited states converge to one while it is zero for the ground state. As the coupling strength increases,  the eigenstates of $H$ involve linear combinations of number states and can be classified according to the eigenvalues of the parity operator. For coupling strength values $g/\omega_{r}\geq 1.5$, the low-lying energy spectrum  of the Hamiltonian $H$ corresponds to a displaced oscillator described by the Schr\"odinger cat states~\cite{Cuiti2010,Nori2010}  with $\ket{\epsilon_0^H}\approx (\mid +, \alpha \rangle + \mid -, -\alpha\rangle)/\sqrt{2}$ and $\ket{\epsilon_1^H}\approx (\mid +, \alpha \rangle - \mid -, -\alpha\rangle)/\sqrt{2}$ ($\alpha=g_\ell/\omega_{r}$ ), where $|\pm \rangle =(|e\rangle \pm |g\rangle)/\sqrt{2}$. These states also contain maximal correlation, as is shown in Fig.~\ref{Fig01}{(b)}.  

Now, we analyze the system and its correlations under thermal equilibrium. Figure \ref{Fig01}~(c) shows the thermal population for the single-qubit Rabi model as a function of $g/\omega_{r}$ for two different temperatures $T=\{T_{c},10T_{c}\}$. We see that in both cases the first two energy states dominate the behavior of the thermal state.  As the coupling strength increases, these two states become asymptotically degenerate which leads to them having essentially the same thermal population for large coupling strength; thus, we can write  ${\rho_{\rm{th}}\approx p_{0}(\ketbra{\epsilon_0^H}{\epsilon_0^H}+ \ketbra{\epsilon_1^H}{\epsilon_1^H})}$ where $p_0=e^{-\beta \epsilon_0^H}/Z$. Figure \ref{Fig01}~(d) shows the quantum correlation for the thermal state as a function of the coupling strength $g/\omega_{r}$ for different temperature values. For small coupling strength ($g/\omega_{r}\approx0.1$) the quantum correlations embedded in the thermal state is close to zero, as $g/\omega_{r}$ increase the correlations increase monotonically until they reach their maximum value, further increasing the coupling strength leads to a monotonically decrease of quantum correlations towards zero. This is because the main contribution comes from the separable ground state $\ket{\epsilon_0^H}=\ket{g,0}$, which holds almost all the thermal population at small values of the coupling strength, and as $g/\omega_{r}$ increases the correlations in the ground state grow until it reaches maximal correlations, similar to the excited states. For large coupling strength values ($g/\omega_{r}\approx1.3$), the thermal state can be approximated to a balanced mixture of cat states, where the embedded correlations go to zero. We observe that the maximal value of the quantum correlations $\mathit{Q}(\rho_{\textrm{Th}})$ occurs at the ultrastrong coupling regime ($0.1\leq g/\omega_{r} \leq 1$) and this maximum does not depend linearly with the temperature. At low temperatures, the thermal state corresponds mainly to the ground state. Consequently, $\mathit{Q}(\rho_{\textrm{Th}})$ follows the same behaviour as the ground state. Nevertheless, as we increase the temperature, the higher excited states start to contribute to the thermal population as depicted in Fig.~\ref{Fig01}~(c), causing a shift in the position of the maximum value of the correlations. To summarize, the degeneration of the energy levels that occurred on the single-qubit Hamiltonian $H$ for significant coupling strength leads to a mixed state of two maximally-correlated quantum states that have zero embedded quantum correlations. Furthermore, the anharmonicity of the spectrum at intermediate coupling strength permits obtaining a thermal state with the maximal embedded quantum correlation, which shifted for a non-trivial contribution of the higher energy levels as the temperature increases.

Next, we extend the study to the MQRS with $N=2$ and $N=3$ qubits. Figure \ref{Fig02}~(a) shows the energy spectrum for $N=2$ (referred to the ground state energy) as a function $g/\omega_{r}$. We observe that the energy of the ground state and first excited state begin to approach each other at smaller values of $g/\omega_{r}$ than when compared with the single-qubit Rabi model.  In Fig.~\ref{Fig02}~(b), we calculate the quantum correlations for the thermal state of the two-qubit Rabi model for temperature above $T_{c}$. We observe similar behavior as in the single-qubit case, but here we can see the maximum value of the quantum correlations has shifted towards smaller values of the coupling strength. We expect analogous behavior for the embedded quantum correlations as we increase the number of qubits. In Fig.~\ref{Fig03}, we plot the quantum correlation of the thermal state for one, two, and three qubits for $T=\{T_{c},5T_{c},10T_{c}\}$. The temperature decreases the quantum correlations in the thermal state, as it drives the state into a  mixed state of maximally correlated states, and beyond a certain temperature the quantum correlations tend to zero. In Fig.~\ref{Fig03}, we can also see that the maximum of the quantum correlations is displaced towards smaller values of $g/\omega_{r}$ as we increase the number of qubits, which displays a collective behavior of the quantum correlations. We could conjecture that the maximal quantum correlations have a power-law dependence on the number of qubits. \par 
\begin{figure}[t]
\centering
\includegraphics[width=0.9\linewidth]{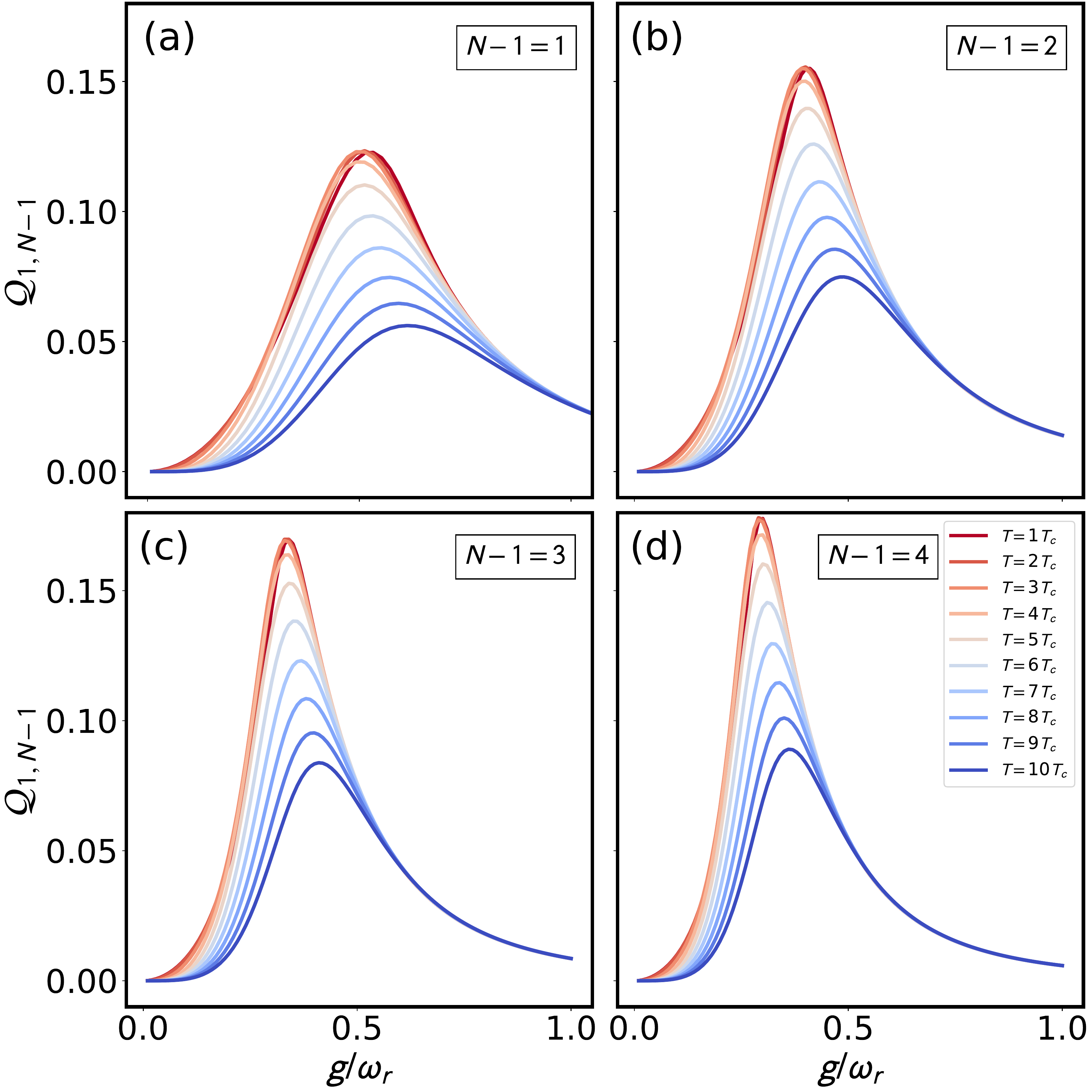}
\caption{The quantum correlations, $\mathit{Q}_{1,N-1}$, of the bipartition composed by the subset of the $N-1$ qubits with one qubit, as a function of the normalized coupling strength $g/\omega_{r}$, considering (a) $N=2$, (b) $N=3$, (c) $N=4$, and (d) $N=5$ qubits. The temperatures for the four cases are indicated in the panel. We have used $\omega_{q,l}=\omega_{r}=\omega$ and $\omega/2\pi=8\,\textrm{GHz}$, with $T_{c}=19\,\textrm{mK}$. }
\label{Fig04}
\end{figure}
Unfortunately, extending our calculations beyond three qubits is too demanding from a computational point of view, since the size of the parameter space increases drastically and quickly becomes a hard computational task~\cite{QDNPHuang}. A strategy to circumvent the impossibility of evaluating the correlation between the many-qubits and the field mode could be the evaluation of quantum correlations among different partitions of the global system. Analyzing the correlations of the partitions as we change the number of qubits can provide hints regarding the exponential dependence of the maxima of correlations and the number of qubits that we have found for a smaller number of qubits. One interesting case arises from the qubits subspace, after tracing the field, by studying how correlations among qubits behave in terms of the total number of qubits. This strategy would allow us to reduce the size of the parameter space drastically, where we only need to optimize on the set of projective measurements on the smallest subspace. 

\begin{figure}[b]
\centering
\includegraphics[width=.9\linewidth]{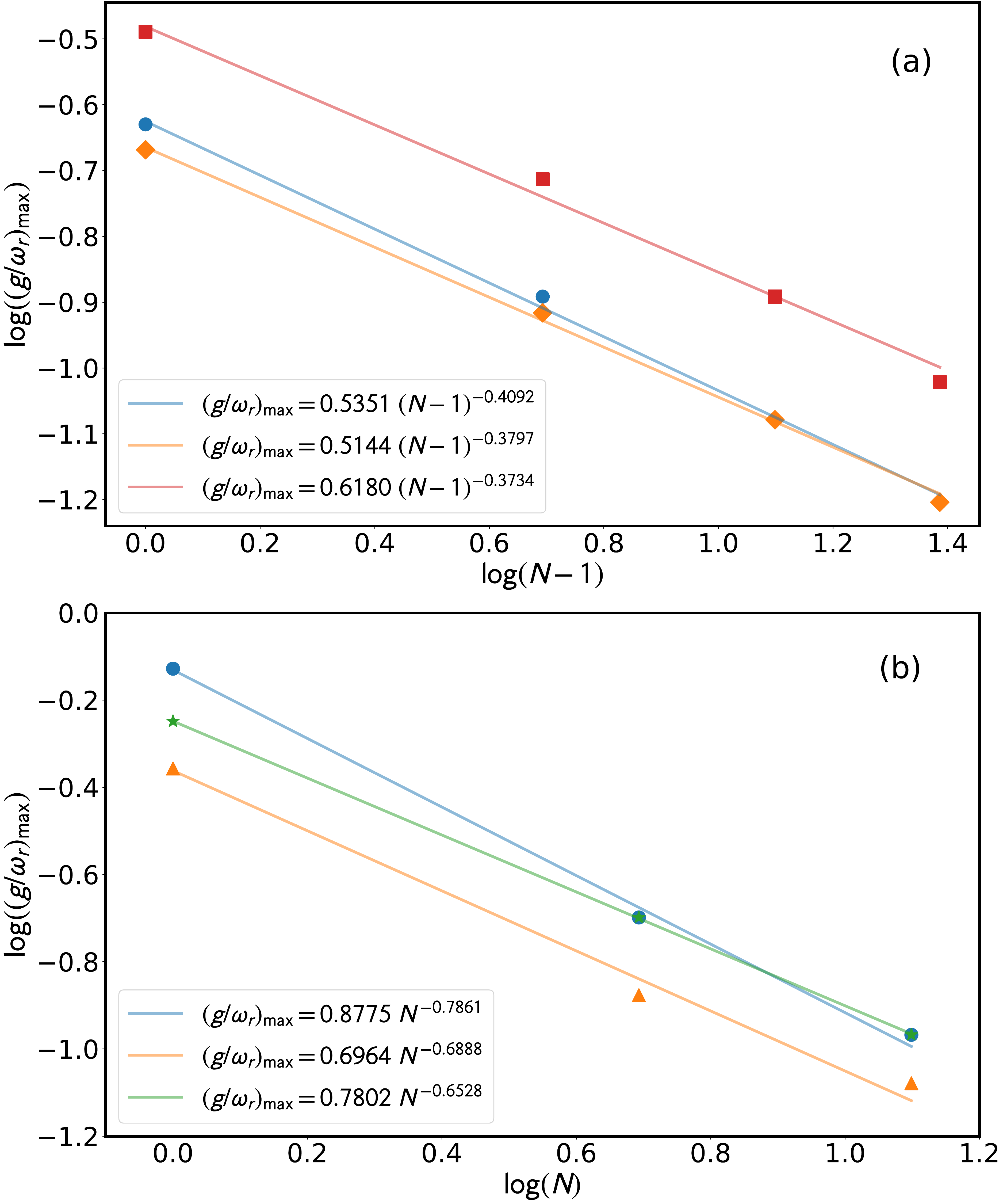}
\caption{(a) The values of the normalized coupling strength $g/\omega_{r}$  at the maximum of the quantum correlations $\mathit{Q}_{1,N-1}$ as a function of the number of qubits in the largest partition, $N-1$, for temperatures  $T=T_{c}$ (blue dots), $T=5T_{c}$ (orange rhoms), and $T=10T_{c}$ (red squares).
The solid lines correspond to curve fittings of the $(g/\omega_{r})_{\max}$ as a function of the $N-1$ qubits. (b) The values of the normalized coupling strength $g/\omega_{r}$ at the maximum of the quantum correlation of the thermal state $\mathit{Q}(\rho_{\textrm{Th}})$ in Figure~\ref{Fig03}, as a function of the number of qubits $N$. The temperature of the thermal state are $T=T_{c}$ (blue dots), $T=5T_{c}$ (orange triangles), and $T=10T_{c}$ (green stars). The solid lines correspond to curve fittings of the $(g/\omega_{r})_{\max}$ as a function of the $N$ qubits.
The power law of each fitting are indicated in the panels. We have used $\omega_{q,l}=\omega_{r}=\omega$ and $\omega/2\pi=8\,\textrm{GHz}$, with $T_{c}=19\,\textrm{mK}$.}
\label{Fig05}
\end{figure}

We define $\rho_{N}={\rm{Tr}_{r}[\rho_{\textrm{Th}}]}$ as the $N$ qubit density matrix. In this scenario, we can analyze the quantum correlations embedded in several bipartitions of $\rho_{N}$. However, as we have assumed symmetric coupling strength ($g_{\ell}=g$), several of these bipartitions will provide the same information. For simplicity and a less demanding calculation, we consider the quantum correlations between the subset formed by $N-1$ qubit with the remaining one, denoted by ${Q}_{N-1,1}$. In Figure~\ref{Fig04}, we depict the quantum correlations ${Q}_{N-1,1}$ as a function of the normalized coupling strength $g/\omega_{r}$ for different temperatures $T$ for $N=2,3,4,5$ qubits. 

We see that the quantum correlations of the reduced density matrix follows the same behavior we have observed before, where the maximum of the quantum correlation is displaced towards smaller values of the coupling strength as we increase the number of qubits. In Fig.~\ref{Fig05}(a), we provide additional evidence about the power-law scaling of the quantum correlation with the number of qubits by plotting $\log(g/\omega_r)_{\textrm{max}}$ against $\log(N-1)$ for three different temperatures, obtaining a correlation coefficient $R^2=0.9978$ for $T=T_c$, $R^2=0.9984$ for $T=5T_c$, and $R^2=0.9929$ for $T=10T_c$. We obtain that the maximum of the quantum correlations $Q_{N-1,1}$ shows an power-law dependence with $N$. Based on these results, we can infer that for any partition of the hybrid multiqubit-field interacting system, quantum correlations among parties exhibit a collective behavior in terms of the number of qubits. In this sense, the power-law dependence in terms of the number of qubits can expected for the multiqubit-field partition as shown in Fig.~\ref{Fig05}~(b), for the thermal state of one, two, and three qubits, respectively, obtaining correlations coefficient $R^2=0.9982$ for $T=T_c$, $R^2=1$ for $T=5T_c$, and $R^2=0.9877$ for $T=10T_c$.

\begin{figure}[t]
\centering
\includegraphics[width=0.9\linewidth]{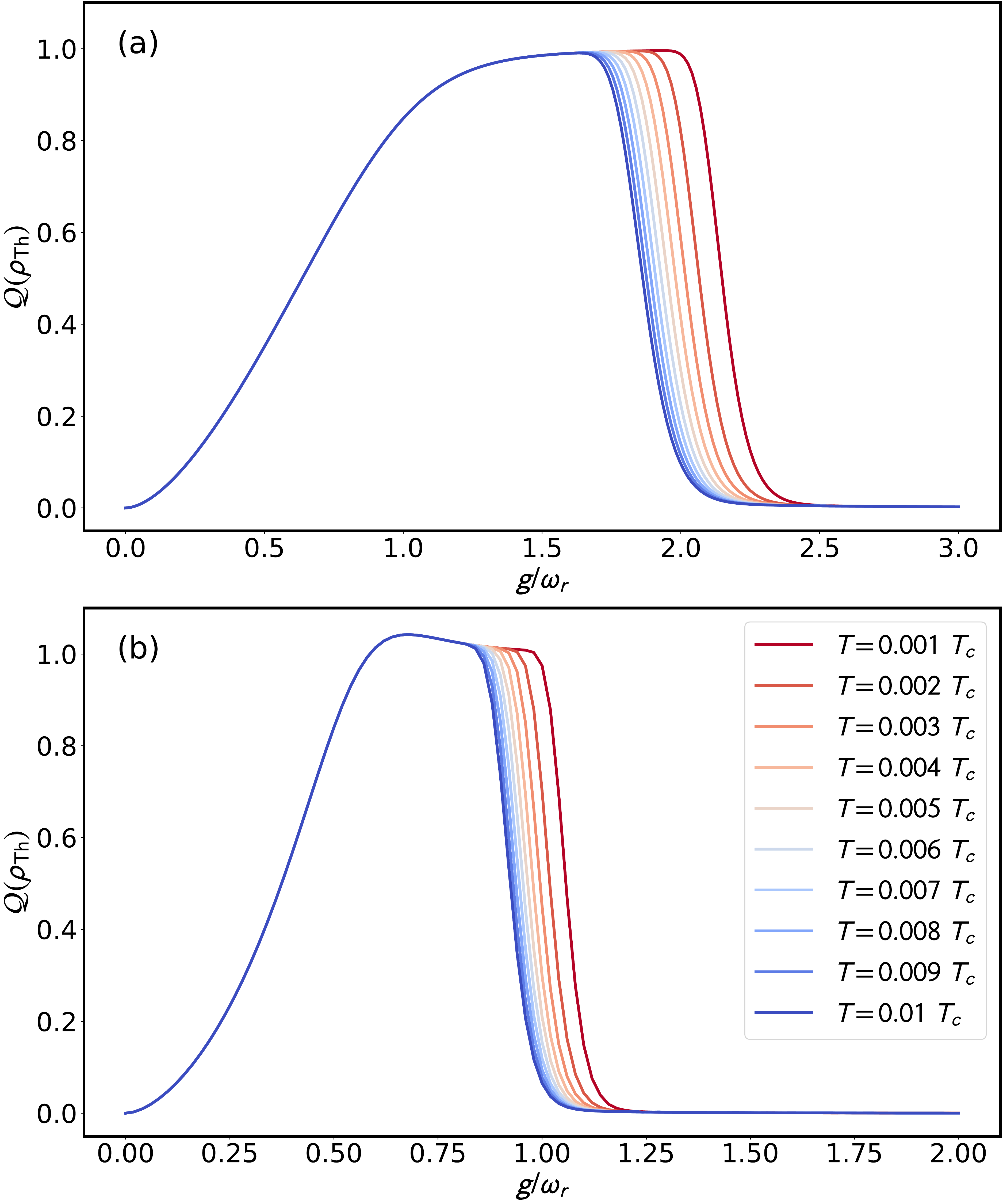}
\caption{Low temperatures regime: Quantum correlation $\mathit{Q}(\rho\textrm{Th})$ in the quantum Rabi model considering (a) $N=1$ and (b) $N=2$ qubits as a function of the normalized coupling strength $g/\omega_{r}$, for several temperatures of the thermal state $T$ indicated in the panels. 
We have used $\omega_{q,l}=\omega_{r}=\omega$ and $\omega/2\pi=8\,\textrm{GHz}$, with $T_{c}=19\,\textrm{mK}$.}
\label{Fig06}
\end{figure}
\begin{figure*}[t]
\centering
\includegraphics[width=0.8\linewidth]{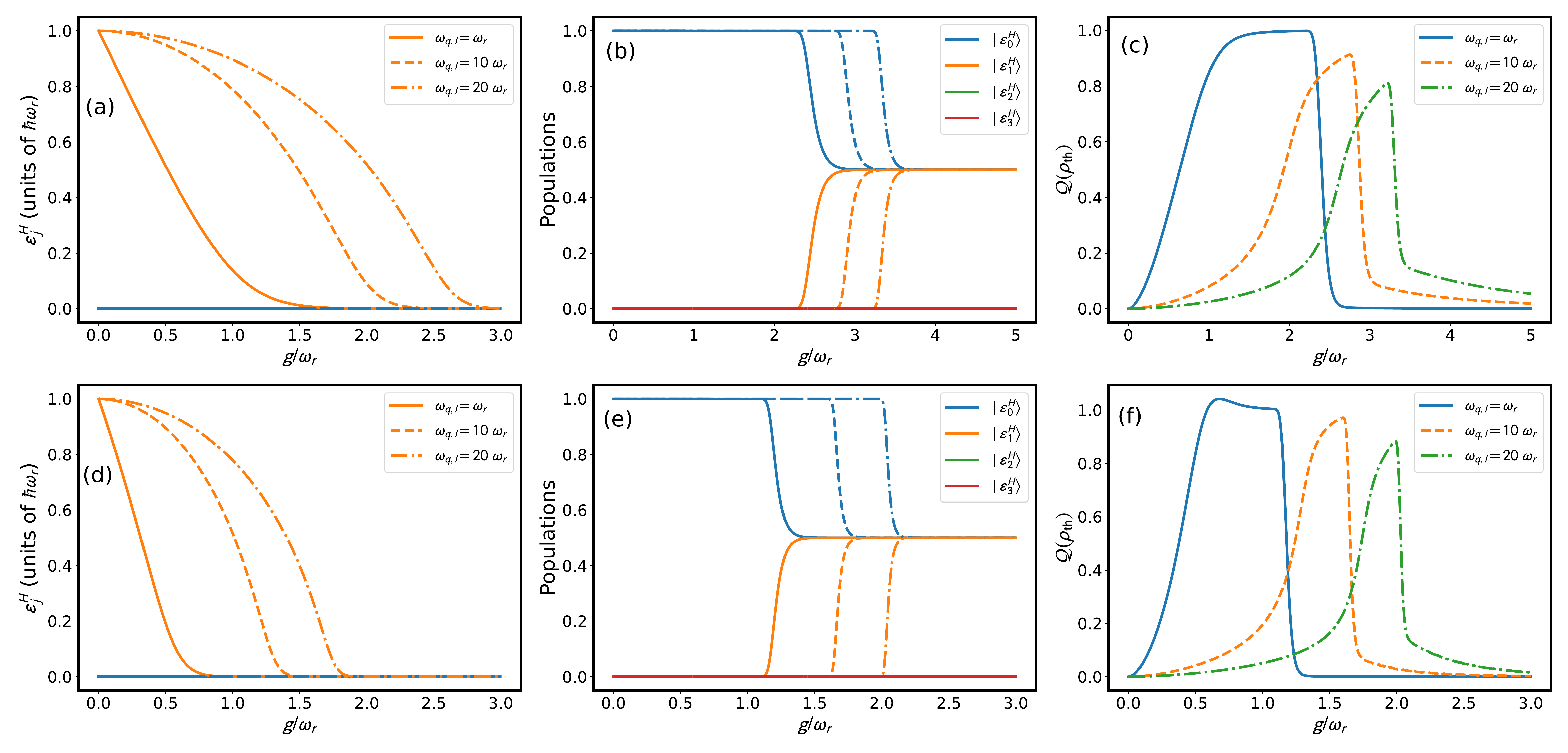}
\caption{Multi-qubit Rabi model considering $N=1$ (upper row) and $N=2$ (lower row) qubits in the low temperature regime, with temperature of the thermal state $T=10^{-4}T_{c}$. 
The qubits frequencies are $\omega_{q,l}=\omega_{r}$ (solid line), $\omega_{q,l}=10\omega_{r}$ (dashed line), and $\omega_{q,l}=20\omega_{r}$ (dash-dotted line). In both cases,  the color lines of  $\ket{\epsilon_j^H}$  are indicated in panels (b) and (e). 
The energy spectrum of the $H$ (left column), the thermal populations (middle column) and, quantum discord of the thermal state $\mathit{Q}(\rho\textrm{th})$ (right column). We consider $\omega_{r} = \omega$ and  $\omega/2\pi=8\,\textrm{GHz}$.}
\label{Fig07}
\end{figure*}
Thus far, we have focused on studying the quantum correlation embedded on a MQRS in a thermal environment above the reference temperature $T_{c}$. It is interesting to investigate now quantum correlations for temperatures below $T_{c}$. 

Figure~\ref{Fig06} shows the correlations for the cases $N=1, 2$ as a function of the coupling strength $g/\omega_{r}$ for different temperatures below $T_c$. We observe in both cases that at intermediate coupling strength values, the correlations are insensitive to the changes in the temperature; this is because the population of the thermal state is mainly concentrated on its ground state, as can be seen in Fig.~\ref{Fig07}~({b}). However, close to the degeneracy point, the correlations tend to split to the right as temperature decreases and eventually vanishes for sufficiently large values of $g/\omega_{r}$. We attribute this effect to the asymptotic degeneracy of the ground state and the first excited state, since for any finite temperature there will be a value of $g/\omega_{r}$ where the energy of both states are close enough to have the same thermal population and the even mixture of these two maximally correlated states results in zero total embedded correlations. The splitting to the right that we see is because smaller temperatures require higher values of $g/\omega_{r}$ for the lowest energy states to have equal population. \par
We also consider an interesting regime that arises in the low-temperature regime when the qubit frequency is significantly larger than the mode frequency ($\omega_{r}\ll\omega_{q}$). In this case, the single-qubit Rabi model experiences a quantum phase transition ($T=0$) at $\eta=2g/\sqrt{\omega_{r}\omega_{q}}=1$~\cite{Hwang2018QPT}. We aim to analyze the quantum correlations embedded on the ground state in this regime as we increase the qubit frequency approaching the limit $\omega_{r}\ll\omega_{q}$. In Fig.~\ref{Fig07}, we plot the energy spectrum, the thermal populations, and the quantum correlation for the single and two-qubit Rabi model as a function of the normalized coupling strength $g/\omega_r$ for different qubit frequencies $\omega_{q,\ell}=\{1,10,20\}\omega_{r}$, respectively. From the figure, we appreciate that these quantities shift towards smaller values of $g/\omega_{r}$ as we consider more qubits, depicted in the upper panel of Fig.~\ref{Fig07}. We can also see that as we increase the qubit frequency the point where the quantum correlations vanish is displaced towards larger values of $g/\omega_{r}$. Since the transition occurs at $\eta=1$, keeping this ratio constant requires stronger values of the coupling strength.

\begin{figure}[t]
\centering
\includegraphics[width=0.9\linewidth]{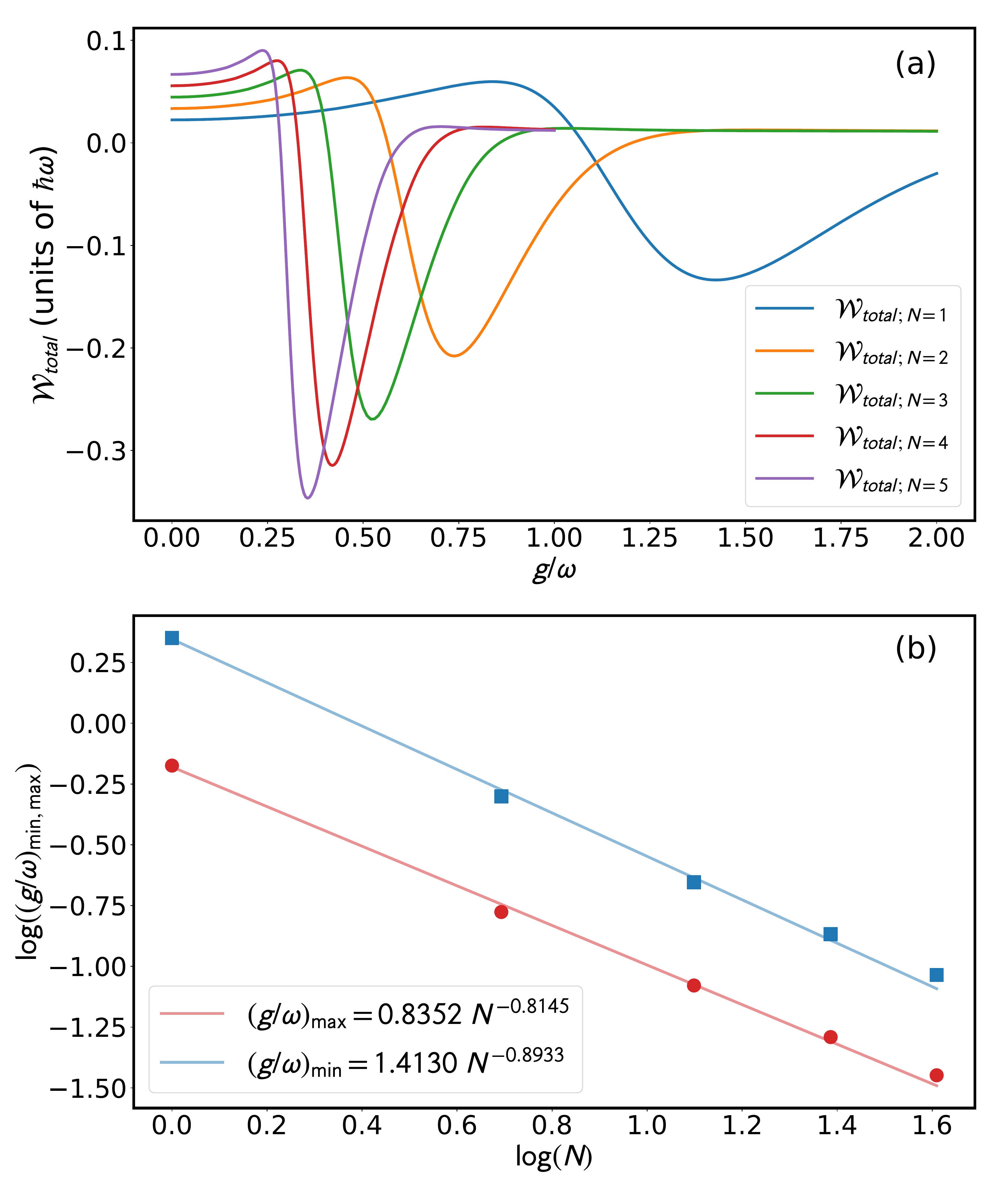}
\caption{(a) The total work extracted in a single cycle of the QHE, $\mathcal{W}_{total}$, as a function of the coupling strength $g/\omega_{r}$ for $N=1,2,3,4,5$ qubits. (b) The values of the coupling strength $g/\omega$ correspond to the maximum (red dots) and minimum (blue squares) of the total work extracted in terms of the number of qubits $N$. The red curve presents a coefficient $R^2=0.9985$ and the blue one $R^2=0.9986$. The red and blue solid lines correspond to curve fittings of the $(g/\omega_{r})_{\max,\min}$ as a function of the number of qubits $N$. The power laws are indicated in the panel.
We have used $\omega_{h}= 2 \omega$, $\omega_{c}=\omega$, and $\omega/2\pi=8\,\textrm{GHz}$, with reservoir temperatures $T_{c}=19\,\textrm{mK}$ and $T_{h}=9T_{c}$.}
\label{Fig08}
\end{figure}

\begin{figure*}[!t]
\centering
\includegraphics[width=0.8\linewidth]{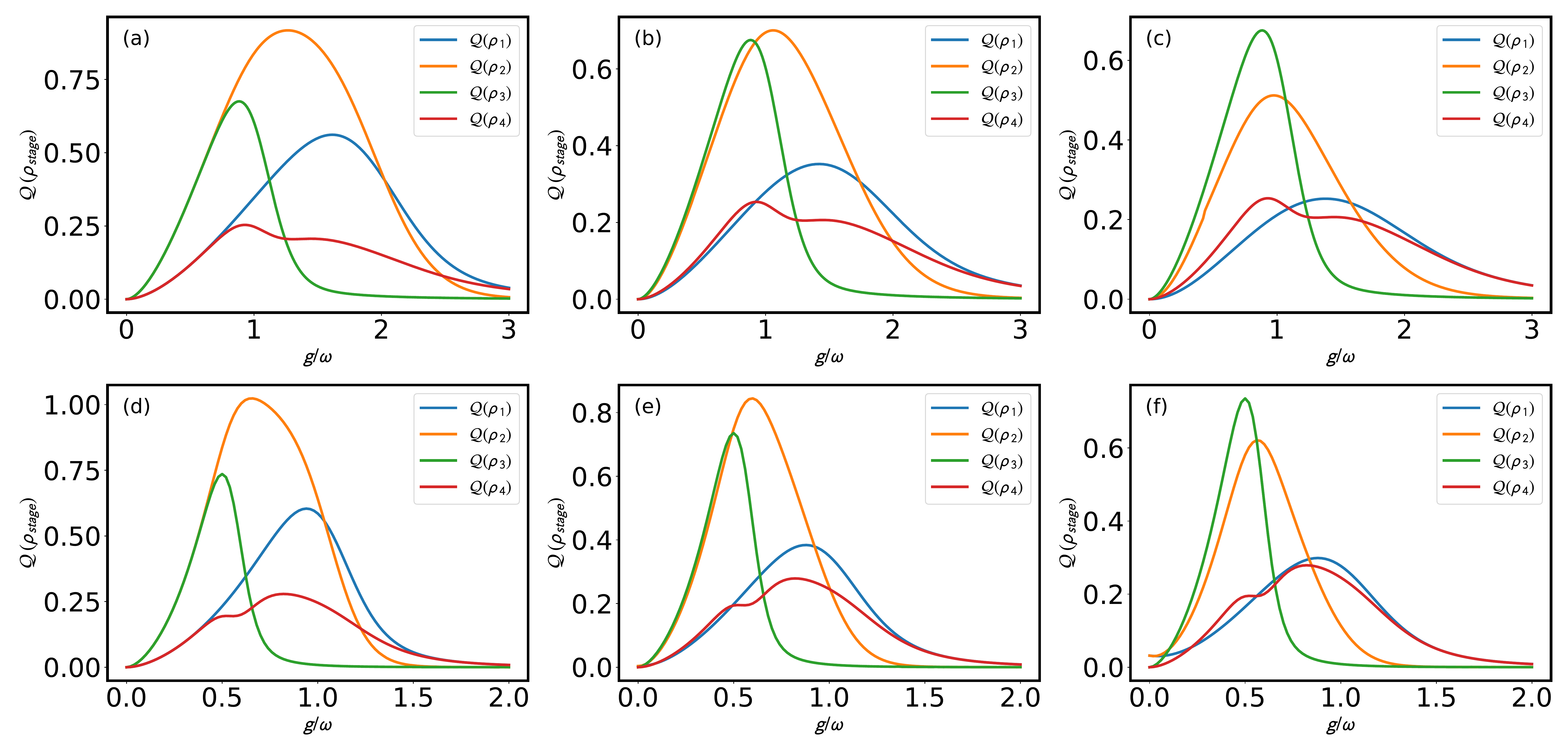}
\caption{QHE in the Otto Cycle: Quantum correlations of the four stages $\mathit{Q}(\rho_{stage})$ as a function of the coupling strength $g/\omega$, considering a working substance of $N=1$ (upper row) and $N=2$ (lower row) qubits interacting with the cavity mode. We considering three different operating temperatures of the hot reservoir $T_{h}=3T_{c}$ (left column), $T_{h}=6T_{c}$ (middle column) and, $T_{h}=9T_{c}$ (right column). 
We have used $\omega_{h}= 2 \omega$, $\omega_{c}=\omega$ with cold reservoir temperature $T_{c}=19\,\textrm{mK}$.}
\label{Fig14}
\end{figure*}

\begin{figure}[!t]
\centering
\includegraphics[width=0.9\linewidth]{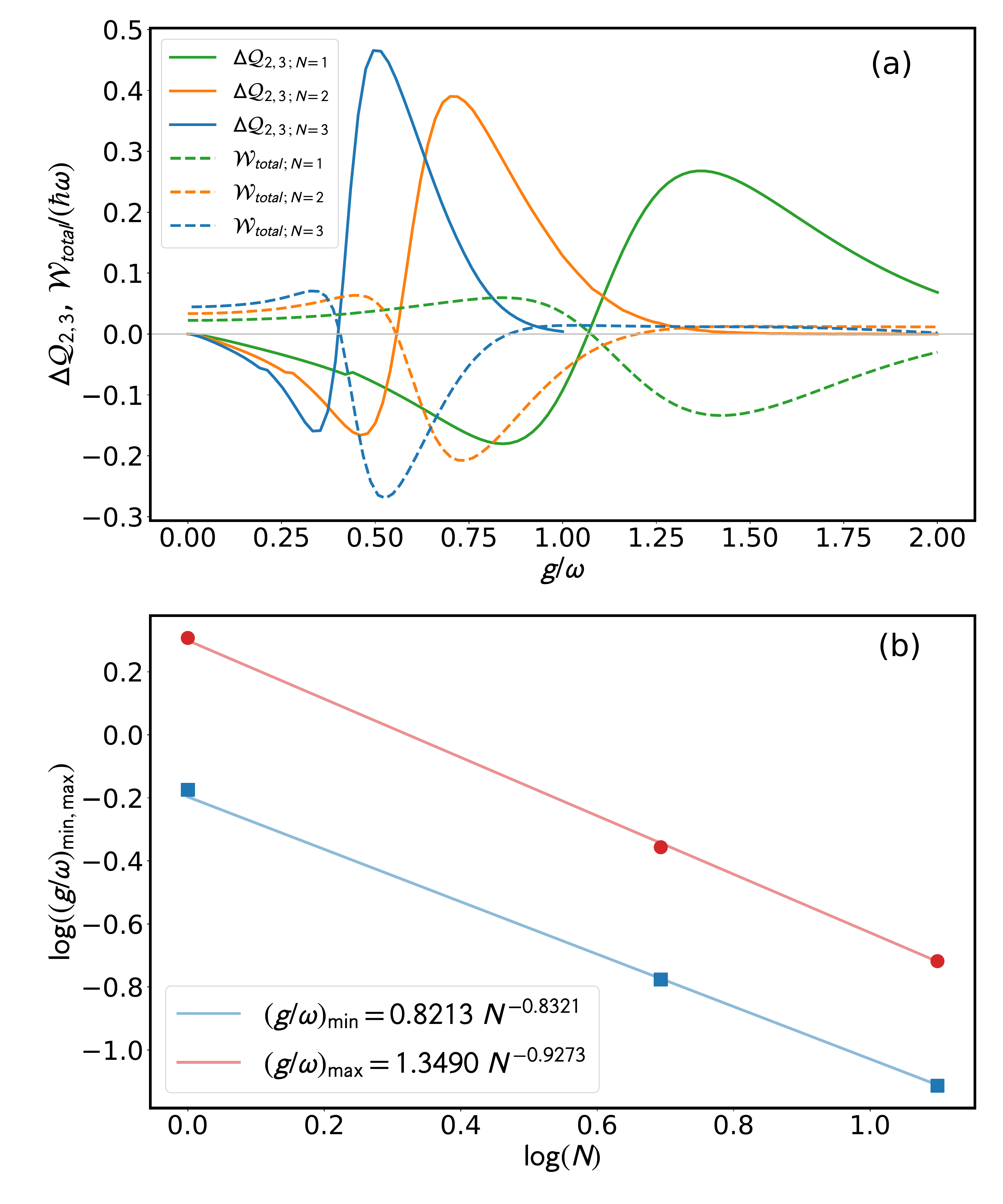}
\caption{(a) Difference of quantum correlations in the cold isochoric stage, $\Delta\mathit{Q}_{2,3}=\mathit{Q}(\rho_{2})-\mathit{Q}(\rho_{3})$ (solid lines), and total work extracted, $\mathcal{W}_{total}$ (dashed lines), both in terms of the coupling strength $g/\omega$ for $N=1,2,3$ qubits. (b) The values of the coupling strength $g/\omega$ corresponding to the maximum (red dots) and minimum (blue squares) of $\Delta\mathit{Q}_{2,3}$ as a function of the number of qubits $N$. The red curve present a coefficient $R^2=0.9996$, and the blue one $R^2=0.9997$. The red and blue solid lines correspond to curve fittings of the $(g/\omega)_{\max,\min}$ as a function of the number of qubits $N$. The power laws are indicated in the panel.
The parameters chosen are $\omega_{h}= 2 \omega$, $\omega_{c}=\omega$, and $\omega/2\pi=8\,\textrm{GHz}$, with reservoir temperatures $T_{c}=19\,\textrm{mK}$ and $T_{h}=9T_{c}$.}
\label{Fig09}
\end{figure}

\section{Ligth-matter quantum correlations on an Otto quantum heat engine}

As an application of our findings, we consider a QHE with a working substance described by Hamiltonian~(\ref{Eq01}) for different numbers of qubits. Given the scaling of quantum correlations, it is interesting to explore the consequences that it can have on the performance of a QHE. We consider a QHE in the Otto cycle~\cite{HenrichEPJST2007,KosloffEntropy2017}, 
characterized by four consecutive processes, they are: the interaction with a hot reservoir, the adiabatic change of a parameter of the system, interaction with a cold reservoir, and the adiabatic restoration of the parameter changed in the second process. We will consider that the coupling strength $g$ will be kept constant throughout the cycle and that the qubits and the cavity are always in resonance $\omega_r = \omega_{q, \ell} = \omega$. The quantum Otto cycle can be described as follows
\begin{enumerate}
	\item Stage 1:  Isochoric process. Thermalization at $T_{h}$. The system, with frequency $\omega = \omega_{h}$ and corresponding Hamiltonian $H_{h}$, thermalizes with an environment at temperature $T_{h}$.  The change in energy in the system is due only to the thermalization process and thus interpreted as heat. 
	\item Stage 2: Adiabatic process. The system is isolated from the thermal reservoir, and its frequency is changed from $\omega_{h}$ to $\omega_{c}$, with $\omega_{h}>\omega_{c}$. This process is performed slow enough to satisfy the quantum adiabatic theorem \cite{NenciuJPA2980} such that the populations of the instantaneous eigenstates remain constant throughout the process. At the end of the process, when the system has frequency $\omega_{c}$ we denote the corresponding Hamiltonian by $H_{c}$.
	\item Stage 3: Isochoric process. Thermalization at $T_{c}$. The system, with frequency $\omega = \omega_{c}$ and Hamiltonian $H_{c}$, thermalizes with an environment at temperature $T_{c}$,  $T_{c}<T_{h}$.  
	\item Stage 4: Adiabatic process. The system is isolated from the thermal reservoir, and its frequency is returned to its initial value, from $\omega_{c}$ to $\omega_{h}$. Satisfying the adiabatic approximation. At the end of the process, the system Hamiltonian is $H_{h}$. \end{enumerate}

During the isochoric process the energy change in the system is interpreted as heat, while the energy change in the adiabatic process is understood as work~\cite{QuanPhysRevE2007}. Since the system thermalizes in our isochoric processes, after completing the cycle, the system returns to its initial configuration. This means that the net energy change during the cycle is zero, which allows us to equate the total work extracted in a single cycle to the total heat exchange between system and environment. Denoting by $Q_{h(c)}$ the heat exchanged during the hot(cold) isochoric stage, we can write\\ 
\begin{eqnarray}
Q_{h} &=& \sum_{n} E_{n}^{h}\big(P_{n}(T_{h}) - P_{n}(T_{c})\big),\\
Q_{c} &=& \sum_{n} E_{n}^{c}\big(P_{n}(T_{c}) - P_{n}(T_{h})\big),\\
W &=& Q_{h} + Q_{c} = \sum_{n} \big( E_{n}^{h} - E_{n}^{c} \big)\big(P_{n}(T_{h}) - P_{n}(T_{c})\big).
\label{Work}
\end{eqnarray}
It must be stressed that this equation is only valid for an idealized quantum Otto cycle where $dQ$ is always zero in the adiabatic process and $dW$ is always zero in the isochoric process.\\ 

First, let us consider the total work extracted in relation to the number of qubits in the working substance. In Fig.~\ref{Fig08}~(a) we show the total work extracted as a function of the coupling $g/\omega$ for different numbers of qubits in the working substance, ranging from $N=1$ to $N=5$. Here, we can see that as we increase the number of qubits, the maximum and minimum values of the total work extracted move towards smaller values of the coupling strength $g/\omega$. We find that the values of $g/\omega$ for the maxima and minima of the total work extracted, denoted by $(g/\omega)_{\textrm{max}}$ and $(g/\omega)_{\textrm{min}}$, respectively, follow a power-law relation with the number of qubits, $N$, which is shown in Fig.~\ref{Fig08}~(b), where the coefficient $R^2=0.9985$ and $R^2=0.9986$ for the red and blue curve respectively. It is interesting to note that this result is reminiscent of Fig.~\ref{Fig05} (a) and Fig.~\ref{Fig05} (b) with similar power-law scalings for maxima and minima of quantum correlations and the number of qubits.

A previous work studied the role of quantum correlations in an Otto cycle with a working substance involving an $N=1$ hybrid light-matter system \cite{Alvarado2017PRA}. In particular, it was stated that the difference in quantum correlations in the hot isochoric stage is indicative of the behavior of the total work extracted. It is interesting to study if this feature prevails as we increase the number of qubits. We have implemented a calculation of quantum correlations for every step in the Otto cycle, for three different operating temperatures, as can be seen in Fig.~\ref{Fig14}, where correlations embedded in steps $1$ to $4$ are shown for $N=1$ and $N=2$ qubits. As we understand from these calculations, for one qubit, in the hot isochoric stage the difference in quantum correlations $\Delta Q_{1,4} \equiv \mathit{Q}(\rho_{1})-\mathit{Q}(\rho_{4})$, decreases as the temperature increases, but they can be well distinguished even for large temperatures. However, for $N=2$, $\Delta Q_{1,4}$ decreases with increasing temperature to the point where distinguishability is practically lost. This vanishing behavior does not carry on to the total work extracted, which means that as we increase the number of qubits $\Delta Q_{1,4}$ is no longer a good indicator of the behavior of the total work extracted. On the contrary, the difference of quantum correlations for the cold isochoric stage $\Delta Q_{2,3} \equiv \mathit{Q}(\rho_{2})-\mathit{Q}(\rho_{3})$ is less distinguishable for lower temperatures compared to the hot isochoric stage, and for larger temperatures the distinguishability of  $\Delta Q_{2,3}$ increases. 

We now study the behavior of the quantum correlations in the cold isochoric stage $\Delta Q_{2,3}$ in relation to the total work extracted, $\mathcal{W}_{\textrm{total}}$ as a function of $g/\omega$. In Fig.~\ref{Fig09} (a), we plot $\Delta Q_{2,3}$ and $\mathcal{W}_{\textrm{total}}$ ($\mathcal{W}_{\textrm{total}}$ should be dimensionless in the plot)  as a function of $g/\omega$ for $N=1,2,3$, we can see that $\Delta Q_{2,3}$ shows the inverse behavior of the total work extracted. Here, the region where $\Delta Q_{2,3}$ attains a minimum, is also where  $\mathcal{W}_{\textrm{total}}$ reaches its maximum value, and thus $\Delta Q_{2,3}$ acts as an indicator of $\mathcal{W}_{\textrm{total}}$ for $N=1,2,3$. Figure~\ref{Fig09} (b) shows that the maximum and minimum values of $\Delta Q_{2,3}$ follow a power law with accurate fittings of $R^2=0.9996$ and $R^2=0.9997$ for the maximums and minimums, respectively. Summarizing, we have found that the difference in quantum correlations in the cold isochoric stage that are indicative of the total work extracted in a quantum Otto heat engine operating with a strongly interacting multiqubit-field hybrid system.

\begin{figure*}[t!]
\centering
\includegraphics[width=0.8\linewidth]{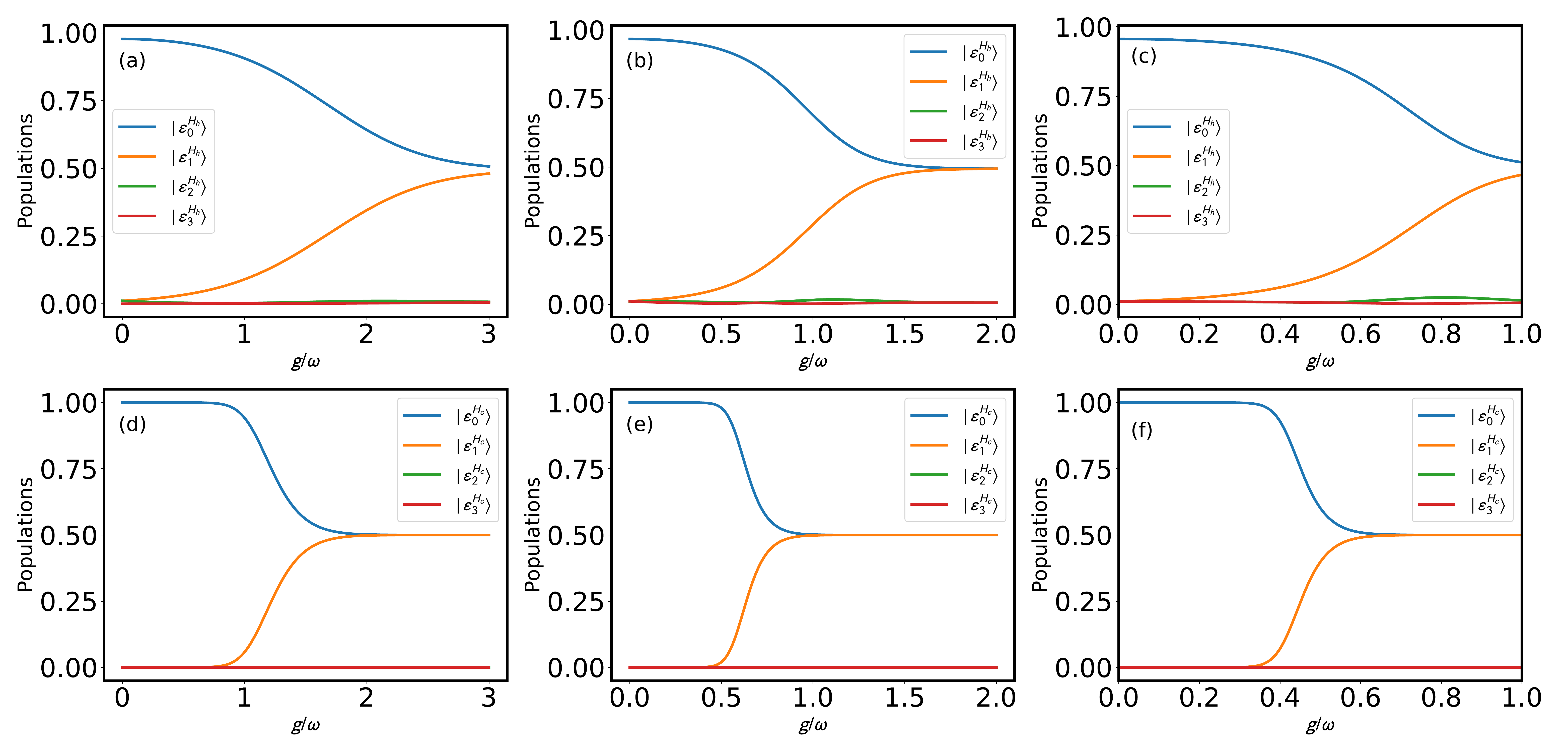}
\caption{QHE in the Otto cycle, involving the working substance interacting with cold and hot reservoirs through four stages.
Upper row: Thermal population of the ground state and first three excited states corresponding to the hot Hamiltonian $H_{h}$, considering a working substance of the (a) $N=1$ (b) $N=2$ and, (c) $N=3$ qubits interacting with the cavity mode.
Lower row: Thermal population of the ground state and three first excited states as a function of the coupling strength $g/\omega$ corresponding to the cold Hamiltonian $H_{c}$, considering a working substance of the (d) $N=1$ (e) $N=2$ and, (f) $N=3$ qubits interacting with the cavity mode.
We have used $\omega_{h}= 2 \omega$, $\omega_{c}=\omega$ with reservoir temperatures $T_{c}=19\,\textrm{mK}$ and $T_{h}=9T_{c}$.}
\label{Fig10}
\end{figure*}

To understand why the quantum correlations in the cold isochoric stage are more descriptive of the work extracted, we analyze the effect of the adiabatic process chosen here on the thermal populations and quantum correlations. In the quantum Rabi model, the ground and first excited state become asymptotically degenerate with increasing coupling $g/\omega$. This means that for any thermal state at finite temperature, $T$, there will be a value of $g/\omega$ where $\epsilon_{1} - \epsilon_{0} \sim k_{B}T$  and the ground and first excited state become almost equally thermally mixed,  as can be seen in Fig.~\ref{Fig07}(b). Now, in our quantum Otto cycle, during the adiabatic processes we change the effective value of $g/\omega$, alternating between $g/\omega_{c}$ and $g/\omega_{h}$ where $\omega_{h} > \omega_{c} \implies g/\omega_{c} > g/\omega_{h}$, and because of asymptotic degeneracy this means $\Delta \epsilon_{1,0}^{h} > \Delta \epsilon_{1,0}^{c}$, where $\Delta \epsilon_{1,0}^{c(h)} = \epsilon_{1}^{c(h)} - \epsilon_{0}^{c(h)} $. Therefore, there can be a region of values of $g/ \omega_{c}$ where during the cycle we may have that for cold isochoric stage $ \Delta \epsilon_{1,0}^{c} \sim k_{B}T_{c} \implies P_{0}^{\textrm{th}}(T_{c}, H_{c}) \sim P_{1}^{\textrm{th}}(T_{c}, H_{c})$, and yet for hot isochoric stage $ \Delta \epsilon_{1,0}^{h} > k_{B}T_{h} \implies P_{0}^{\textrm{th}}(T_{h}, H_{h}) > P_{1}^{\textrm{th}}(T_{h}, H_{h})$, therefore, the cold thermal state will have more thermal mixing than the hot thermal state, since it is enhanced by degeneracy as we can see in Fig. \ref{Fig10}, where we show the population of the different energy levels as a function of the coupling strength for the thermal state. Now, because the adiabatic process preserves the populations, when we do the adiabatic process $\rho_{3} = \rho_{\textrm{th}}(T_{c}, H_{c}) \rightarrow \rho_{4}$ if we have a cold thermal state with degeneracy enhanced mixing, the hot isochoric process will actually partially unmix the state, creating quantum correlations in the process.  

\begin{figure*}[t!]
\centering
\includegraphics[width=0.8\linewidth]{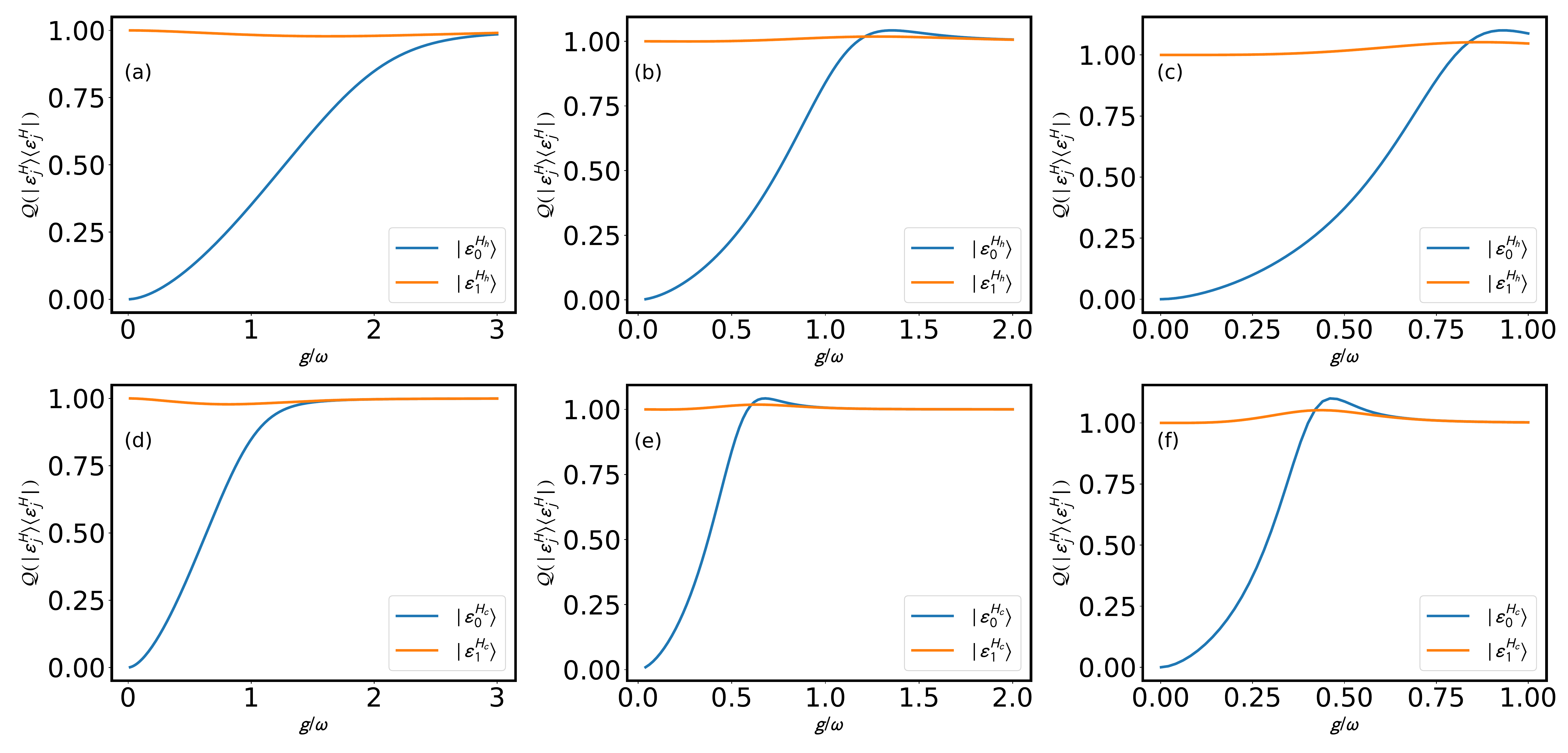}
\caption{Quantum correlations of the ground $\mathit{Q}(\vert\epsilon_{0}^{H}\rangle\,\langle\epsilon_{0}^{H}\vert)$ and the first excited states $\mathit{Q}(\vert\epsilon_{1}^{H}\rangle\,\langle\epsilon_{1}^{H}\vert)$  as a function of the coupling strength $g/\omega$ for (upper row) the hot and (lower row) cold Hamiltonian, considering (left column) $N=1$, (middle column) $N=2$ and (right column) $N=3$ qubits interacting with the cavity mode. We have used $\omega_{h}= 2 \omega$, $\omega_{c}=\omega$ with reservoir temperatures $T_{c}=19\,\textrm{mK}$ and $T_{h}=9T_{c}$}
\label{Fig11}
\end{figure*}

On the other hand, the quantum correlations of the ground state, $\mathit{Q} (\vert\epsilon_{0}^{H} \rangle\,\langle \epsilon_{0}^{H}\vert)$, and the first excited states, $\mathit{Q}(\vert\epsilon_{1}^{H}\rangle \,\langle \epsilon_{1}^{H}\vert)$, play an important role in the behavior of $\Delta Q_{1, 4}$ and $\Delta Q_{2, 3}$. In Fig.~\ref{Fig11}, we show the  $\mathit{Q} (\vert\epsilon_{0}^{H} \rangle\,\langle \epsilon_{0}^{H}\vert)$ and  $\mathit{Q}(\vert\epsilon_{1}^{H}\rangle \,\langle \epsilon_{1}^{H}\vert)$  as a function of $g/\omega$, we can see that the excited state is always in an entangled state whereas the ground state starts with $0$ correlations when $g/\omega \sim 0$ and eventually reaches maximally entangled state as $g/\omega$ increases. Notice that the states become approximately degenerate in the same region where their quantum correlations become similar. Since for the cold thermal state, the complete mixing occurs for large values of $g/\omega_{c}$, where both the ground state and the excited states are maximally entangled states, their equal mix results in zero quantum correlations. A special situation occurs when the degeneracy makes the cold thermal state to be more mixed than the hot thermal state, but the ground state in the hot Hamiltonian has not yet reached the same amount of correlations as its excited state. This is important because of state $\rho_{4}$ which combines the thermal populations from the cold thermal state, and the eigenstates of the hot Hamiltonian. In this situation, as we increase the coupling $g/\omega_{h}$, in $\rho_{4}$ appears a competition between the degeneracy induced mixing which decreases the quantum correlations of $\rho_{4}$, and the quantum correlations of the ground state that increase with $g/\omega_{h}$ in this region. This competition ends distorting the ability of the correlation difference $\Delta Q_{1,4}$ to indicate the behavior of the total work extracted. On the other hand, $\Delta Q_{2,3}$ never has this issue and its behavior is more stable, which makes it a better indicator of the total work extracted.

Concerning the calculation leading to Fig.~\ref{Fig14}, we have considered a dimensional reduction of the $N$-qubits Hilbert space. As can be easily understood from Eq. (\ref{Eq01}), for homogeneous coupling of qubits to the field mode, the ground and first excited state of the system resides on the symmetric multiplet of the qubits Hilbert space. Expressing the states belonging to angular momentum basis, for two qubits the singlet corresponding to $j=0$ is a dark state with energy $E=\hbar \omega_r n$ independent of the coupling strength. The ground and first excited states are combinations of even states involving symmetric states of $j=1$ triplet. For three qubits, states belonging to $j=1/2$ doublets are excited states, and the ground and first excited state is a combination of even states involving the states belonging to the $j=3/2$ multiplet. Summarizing,  the calculation of correlations can be carried out in reduced Hibert space, namely, $d=3$ for two qubits and $d=4$ for 3 qubits, instead of $d=4$ and $d=8$ respectively. This dimensional reduction allows a drastic reduction in the number of parameters needed for the optimization problem.

\section{Conclusion}

In this work, we have studied the quantum correlations embedded in a thermal state of a hybrid light-matter system described by the multi-qubit Quantum Rabi model. Specifically, we studied the behavior of the quantum correlations, given by the quantum discord, as a function of the temperature, the coupling strength, and the number of qubits. We found that in thermal equilibrium, the correlations exhibit a maximum for a given coupling strength; as the number of qubits increases, the coupling strength that attains the maximum value of correlations becomes smaller, obtaining numerical evidence of a power-law dependence. 

We found that the quantum correlations of the thermal state are dominated by the correlations in the ground state, which grow from zero at low coupling strength and monotonously increase until a highly correlated state at sufficiently high coupling strength. For $N=1$, we found that the ground state and excited state reach an effective maximally correlated state between the qubit and the single quantum field mode. In the thermal state, the correlations are close to zero at low coupling strength; as the coupling strength increases, the correlations reach a maximum value. If the coupling strength increases past this point, the quantum correlations asymptotically decrease toward zero. The local maximum in the quantum correlations of the thermal state is a consequence of the competition between the increase in correlations in the ground state and the thermal mixing that is enhanced by the degeneracy in the energy spectrum of the quantum Rabi model. The asymptotic degeneracy of the spectrum enhances the mixing of eigenstates in thermal equilibrium, and the quantum correlations tend to zero as the coupling strength increases. As a consequence, even for very small finite temperatures, there is always a sufficiently high coupling strength for which the thermal state has zero correlations. 

 We have calculated the quantum discord for the set of one, two, and three qubits with the single quantum field mode. As the optimization process is computationally demanding, we have considered up to six qubits where the calculation of the quantum correlations considers a bipartition between one of the qubits and the rest $N-1$, by tracing the quantum field mode. In this case, we obtained similar power-law-like dependence for the quantum correlations, suggesting that the correlations embedded in the set of qubits represent the whole system. 

Quantum correlations in thermal states have potential applications in emerging quantum technologies, such as in quantum thermodynamics. We have considered as a possible application of our study, a thermal machine using the quantum Otto cycle, with a working substance consisting of a hybrid light-matter system described by the multi-qubit Rabi model. We have found that the difference in quantum correlations in the cold isochoric stage is a good indicator of the operating regime and total work extracted in a quantum Otto heat engine operating with a strongly interacting multiqubit-field hybrid system. This is because in our model when the degeneracy causes the cold thermal state to be more mixed than the hot thermal state, the hot isochoric stage tends to unmix the state, negatively impacting the difference of quantum correlations. On the other hand, the cold isochoric stage is not affected by degeneracy enhanced mixing and can perform as a good indicator of total work extracted for cases with an increasing number of qubits. 

\section*{Acknowledgments}
We acknowledge financial support from Vicerrector\'ia de Investigación, Desarrollo e Innovaci\'on USACH POSTDOC\_DICYT 042131RA\_POSTDOC, ANID Subvenci\'on a la Instalaci\'on en la Academia SA77210018, and Financiamiento Basal para Centros Científicos y Tecnológicos de Excelencia (Grant No. AFB 220001). F. A. C. L acknowledges support from the project AIDAS: AI, Data Analytics and Scalable Simulations, a joint virtual laboratory gathering the
Forschungszentrum Jülich (FZJ) and the French Alternative Energies and Atomic Energy Commission (CEA).

\end{document}